\documentclass{elsart}
\usepackage{epsf}
\usepackage{epsfig} 
\begin{document}
\begin{frontmatter}
\baselineskip=12pt
\title{On the Gravitomagnetic Time Delay}
\author[ic]{I. Ciufolini}, \author[sk]{S. Kopeikin}, \author[sk]{B. Mashhoon\corauthref{cor1}}\ead{mashhoonb@missouri.edu}, \author[fr]{F. Ricci}
\address[ic]{Dipartimento di Ingegneria dell'Innovazione, Universit\`a di Lecce, Via Monteroni, 73100 \indent Lecce, Italy}
\address[sk]{Department of Physics and Astronomy, University of Missouri-Columbia, 
Columbia, \indent Missouri 65211, USA}
\address[fr]{Dipartimento di Fisica, Universit\`a di Roma, `La Sapienza', 
Piazzale Aldo Moro
5, \indent 00185 Roma, Italy}
\corauth[cor1]{Corresponding author}

\begin{abstract}
We study the gravitational time delay in ray propagation due to rotating masses in the
linear approximation of general relativity.  Simple expressions are given 
for the
gravitomagnetic time delay that occurs when rays of radiation cross a slowly rotating
shell, equation (14), and propagate in the field of a distant rotating 
source,
equation (16).  Moreover, we calculate the {\it local} gravitational time 
delay in the
G\"odel universe.  The observational consequences of these results in the case of weak
gravitational lensing are discussed.
\end{abstract}

\begin{keyword}
gravitomagnetism \sep time delay \sep gravitational lensing
\PACS 04.20.Cv; 95.30.S
\end{keyword}
\maketitle
\end{frontmatter}

\section{Introduction}\label{sec:1} 

Consider the gravitational field of astronomical sources in the linear 
approximation of
general relativity.  The spacetime metric may be expressed in the 
quasi-inertial
coordinates $x^{\mu} = (ct,{\bf x})$ as
$g_{\mu\nu}=\eta_{\mu\nu} + h_{\mu\nu},$ where $\eta_{\mu\nu}$ is the 
Minkowski metric
with signature +2.  In the absence of perturbing potentials 
$h_{\mu\nu}(x)$, rays of
electromagnetic radiation propagate along straight lines defined by 
$dx^i/dt = c{\hat
k}^i$, where ${\bf\hat k}$ is the constant unit propagation vector of the 
signal. These zeroth-order null geodesics will be employed throughout this work.  In
the exterior gravitational field of astronomical sources, however, the 
rays bend due to
the attraction of gravity as they follow geodesics of the spacetime 
manifold that are
null. It turns out that in order to evaluate the gravitational time delay to first order in $h_{\mu\nu}$ only the null condition is required, i.e.
\begin{equation} g_{\mu\nu}\;dx^{\mu}\;dx^{\nu} = 0\;\;.
\end{equation}
We are interested in the consequences of this condition
for the propagation of a ray on a background global {\it inertial
frame}.  Therefore, it follows from equation (1) that $c^2dt^2 -
|d{\bf x}|^2 = h_{\mu\nu}\;dx^{\mu}\;dx^{\nu}\;,$ where $|d{\bf
x}|^2 = \delta_{ij}\;dx^i\;dx^j$.  To first order in the
perturbation, the bending of the ray may be neglected in
evaluating the right-hand side of this result, which may be
written as $c^2\;h_{\alpha\beta}\;k^{\alpha}\;k^{\beta}\;dt^2.$
Here $k^{\alpha} = (1, {\bf\hat k})$ is such that
$\eta_{\mu\nu}\;k^{\mu}\;k^{\nu} = 0$.  One may therefore write
equation (1) in the form
\begin{equation} cdt = \left( 1 +
\frac{1}{2}h_{\alpha\beta}\;k^{\alpha}\;k^{\beta}\right)\;|d{\bf x}|\;\;.
\end{equation}
Let the ray propagate from a point $P_1 : (ct_1, {\bf x}_1)$ to 
a point $P_2:
(ct_2, {\bf x}_2)$ in the background inertial frame; then,
\begin{equation} t_2 - t_1 = \frac{1}{c}|{\bf x}_2 - {\bf x}_1 | +
\frac{1}{2c}\:k^{\alpha}\:k^{\beta}\int^{P_2}_{P_1}h_{\alpha\beta}(x)\;dl\;\;,
\end{equation}
where $dl = |d{\bf x}|$ denotes the Euclidean length element 
along the
straight line that joins $P_1$ to $P_2$.  It follows from equation (3) 
that the {\it
gravitational time delay} $\Delta$ is given by
\begin{equation}
\Delta_G =
\frac{1}{2c}\int^{P_2}_{P_1}\;h_{\alpha\beta}(x)\;k^{\alpha}\;k^{\beta}\;dl\;\;.
\end{equation}
Note that in this expression $h_{\alpha\beta}$ may be replaced 
by
${\bar h}_{\alpha\beta} = h_{\alpha\beta} - 
\frac{1}{2}\eta_{\alpha\beta}h,$ since
$\eta_{\alpha\beta}k^{\alpha}\:k^{\beta}=0$.  Here $h = {\rm tr} 
(h_{\mu\nu})$.
Moreover, under a gauge transformation of the gravitational potential
$h_{\mu\nu}\rightarrow h_{\mu\nu} + \epsilon_{\mu,\nu} +
\epsilon_{\nu,\mu}$, corresponding to an infinitesimal coordinate 
transformation
$x^{\mu}\rightarrow\;x^{\mu}-\epsilon^{\mu}$, equation (4) simply gives 
the time delay
in the new coordinates.

In physical circumstances where the linearized Einstein equations are applicable, the
time delay $\Delta_G$ may be combined with the proper time
$\tau$ measured by an observer moving with velocity ${\bf v} =
c\mbox{\boldmath$\beta$}$,
\begin{equation} 
d\tau = cdt\;\sqrt{-g_{00}-2g_{oi}\beta^i - 
g_{ij}\beta^i\beta^j}\;\;,
\end{equation}
in order to arrive at physically measurable predictions of the 
theory.

The gravitational time delay was studied by Shapiro \cite{[1]} in terms
of radar echo delay and subsequently by a number of investigators
(see \cite{[2]}--\cite{add-ref2} and references cited therein).  The general derivation
of the linear effect (4) given here is free of special
assumptions and can be employed in most situations of physical
interest.

The gravitomagnetic time delay is discussed in sections 2 and 3.
The G\"{o}del universe is considered in section 4; we study the
local gravitational time delay in this rotating universe. In
section 5 we analyze the gravitomagnetic time delay in different
images of the same source due to gravitational lensing. Finally,
section 6 contains a brief discussion of our
results.

\section{Gravitoelectromagnetism}\label{sec:2}

Let us now consider the exterior field of slowly moving sources such that 
the
trace-reversed potential ${\bar h}_{\mu\nu}$ satisfies $\Box\:{\bar 
h}_{\mu\nu}=-(16\pi
G/c^4)\: T_{\mu\nu}$, once the gauge condition
${\bar h}^{\mu\nu}\,_{,\nu}=0$ has been imposed.  We are interested in the 
particular
retarded solution
\begin{equation} {\bar h}_{\mu\nu} = \frac{4G}{c^4}\int 
\frac{T_{\mu\nu}(ct - |{\bf x} -
{\bf x}^{\prime}|\;,\;{\bf x}^{\prime})}{|{\bf x} - {\bf 
x}^{\prime}|}\;d^3{\bf x}^{\prime}\;\;,
\end{equation}
which can be used directly in equation (4) to compute the 
gravitational time
delay.  The generalization of the Shapiro effect to time-dependent 
situations has been
the subject of a recent investigation \cite{[4]}.

We confine our study here to the {\it stationary} field of a slowly 
rotating
astronomical body such that ${\bar h}_{00} = 4\Phi_g/c^2, {\bar h}_{0i} = 
-2({\bf
A}_g)_i/c^2$ and ${\bar h}_{ij} = O(c^{-4}),$ where $\Phi_g(x)$ is the 
gravitoelectric
potential, ${\bf A}_g(x)$ is the gravitomagnetic vector potential
$(\mbox{\boldmath$\nabla$}\cdot{\bf A}_g = 0)$ and we neglect all terms of 
order
$c^{-4}$ including the spatial potentials ${\bar h}_{ij}$.  Far from the 
source
\begin{equation}
\Phi_g \sim\frac{GM}{r}\;\;,\;\;{\bf A}_g\sim\frac{G}{c}\;\frac{{\bf 
J}\times{\bf
r}}{r^3}\;\;,
\end{equation}
where $M$ and $J$ are the total mass and angular momentum of the 
source.  We
define the gravitoelectric field to be ${\bf E}_g 
=-\mbox{\boldmath$\nabla$}\Phi_g$ and
the gravitomagnetic field to be ${\bf B}_g = 
\mbox{\boldmath$\nabla$}\times{\bf
A}_g$.  In developing this analogy with electrodynamics, one encounters 
extra
numerical factors that cannot be avoided.  For instance, the field 
equation for the
standard gravitomagnetic vector potential has an extra factor of $-4$ in 
comparison
with Maxwell's theory [2].  In this connection, we use the following 
convention that is
consistent with the gravitational Larmor theorem \cite{[6]}:  The standard 
formulas of
classical electrodynamics are applicable, except that for a source of inertial mass
$M$, the gravitoelectric charge is $M$ and the gravitomagnetic charge is 
$2M$ (in units
such that $G = 1$).  On the other hand, for a test particle of mass
$m$, the gravitoelectric charge is $-m$ and the gravitomagnetic charge is
$-2m$.  The source and the test particle have opposite charges to ensure 
that gravity
is attractive.  Moreover, the ratio of the gravitomagnetic charge to the
gravitoelectric charge is always 2, since linearized gravity is a spin-2 
field.

We find from equation (4) that $\Delta_G = \Delta_{GE} + \Delta_{GM}$, 
where
\begin{equation}
\Delta_{GE} = \frac{2}{c^3}\;\int^{P_2}_{P_1}\;\Phi_g\;dl
\end{equation}
is the {\it Shapiro time delay} and
\begin{equation}
\Delta_{GM} = -\frac{2}{c^3}\;\int^{P_2}_{P_1}\;{\bf A}_g\cdot d{\bf x}
\end{equation}
is the {\it gravitomagnetic time delay}, which is simply 
proportional to the
line integral of the gravitomagnetic vector potential.

Let us now suppose that we can arrange via ``mirrors'' --- these could be 
transponders
on spacecraft --- to have the rays travel on a closed trajectory as in 
Figure 1.  Then
the net gravitomagnetic time delay for a closed loop in the positive sense 
(i.e.
counterclockwise) is
\begin{equation}
\Delta^+_{GM} = -\frac{2}{c^3}\;\oint{\bf A}_g\cdot d{\bf x} = 
-\frac{2}{c^3}\int{\bf
B_g}\cdot d{\bf S}\;\;,
\end{equation}
which is simply proportional to the gravitomagnetic flux 
threaded by a
surface $S$ whose boundary is the closed loop under consideration \footnote{ It is interesting to note that
the
direct echo delay (i.e. when the loop degenerates to a line) vanishes in the gravitomagnetic case as the effective 
area is
zero.}.  Imagine now a point $P$ on the trajectory (cf. Figure 1).  If the 
rays travel
along the same path but in the opposite direction, the net gravitoelectric 
time delay
measured at $P$ will be the same while the gravitomagnetic time delay will 
change
sign, $\Delta^{-}_{GM} = -\Delta^+_{GM}$.  Thus the total time difference 
at $P$ for
the rays to go around the loop in opposite directions is given by
\begin{equation}\label{11} 
t_+ - t_- = -\frac{4}{c^3} \oint {\bf A}_g\cdot d{\bf x} 
=
-\frac{4}{c^3}\int {\bf B_g}\cdot d{\bf S}\;\;.
\end{equation}
A different derivation of this result based on the propagation 
of
electromagnetic radiation in the gravitational field of a rotating mass is 
contained in ref.
\cite{[7]} \footnote{A sign error in this reference must be corrected:  In ref. \cite{[7]}, 
the relation
$\Phi_+ - \Phi_- =\omega(t_+ - t_-)$ contains an errant minus sign.  
Therefore in
equations (20), (21) and related discussion in section 3 of \cite{[7]}, the sign of 
$t_+-t_-$ has to be reversed.}.

Let us now estimate this time difference for the case of GPS
signals traveling around the Earth.  We find that $t_+ -
t_-\approx -8\pi GJ_{\oplus}/(c^4 R_{\oplus})\approx-10^{-16}$
sec.  By contrast, the Shapiro time delay amounts to $2\times
10^{-10}$ sec for clocks in GPS orbit \cite{[8]}, \cite{[9]}, so that this
gravitoelectric effect may be measurable in the near future.
In such an experiment, the time of flight of the signal is monitored 
and eventually measured by a 
control station on the Earth \cite{[8]}, \cite{[9]}. 
The
corresponding gravitomagnetic effect is about two million times
smaller and can be completely neglected at present.  Other very
small angular momentum effects in laboratory-based optical
interferometry experiments have been studied in \cite{[10]}, \cite{[11]}.

\section{Gravitomagnetic time delay}\label{sec:3}

In this section, we provide explicit formulas for $\Delta_{GM}$ for the 
case of
radiation crossing a slowly rotating thin shell of matter as well as 
propagating in the
exterior field of a distant rotating mass.

In electrodynamics, the magnetic vector potential inside a
spherical shell of uniform charge density with total charge $Q$
and radius $R_0$ rotating with constant frequency ${\omega}$ is
given by ${\bf A}=\frac{1}{2}{\bf B}\times{\bf r}$, where ${\bf B}
=\frac{2}{3}Q\mbox{\boldmath$\omega$}/(cR_0)$ is the uniform
magnetic field inside the shell.  The corresponding electric
field vanishes.  Therefore, the gravitomagnetic vector potential
inside a uniform shell of mass $M$, radius $R_0$ and rotational
frequency $\mbox{\boldmath$\omega$}$ is \cite{[12]}, \cite{[13]}
\begin{equation} {\bf A}_g 
=\frac{2GM}{3cR_0}\;\mbox{\boldmath$\omega$}\times{\bf
r}\;\;,
\end{equation}
where the center of the shell is the spatial origin of
a background inertial frame.  Thus inside the shell ${\bf B}_g =
4GM\mbox{\boldmath$\omega$}/(3cR_0)$ and ${\bf E}_g=0$.  The
gravitomagnetic time delay is therefore
\begin{equation}
\Delta_{GM}=-\frac{4GM}{3c^4R_0}\;\mbox{\boldmath$\omega$}\cdot\int^{P_2}_{P_1}{\bf
r}\times d{\bf r}\;\;.
\end{equation}
Let $P_1$ and $P_2$ be the two points on the spherical shell 
indicating the
points at which the ray enters and leaves the sphere, respectively.  Then 
a simple
calculation shows that
\begin{equation}\label{14}
\Delta_{GM}=-\frac{4GMR_0}{3c^4}\;\mbox{\boldmath$\omega$}\cdot({\bf\hat 
r}_1\times
{\bf\hat r}_2)\;\;,
\end{equation}
where ${\bf\hat r}_1$ and ${\bf\hat r}_2$ are unit vectors 
indicating the
positions of $P_1$ and $P_2$, respectively, from the center of the 
sphere.  Note that
the result vanishes if $\mbox{\boldmath$\omega$}$, ${\bf r}_1$ and ${\bf 
r}_2$ are
in the same plane; in particular, $\Delta_{GM} = 0$ if the ray passes through the
center of the spherical shell.

Let us next consider the exterior field of a rotating mass with its center 
of mass at
the origin of spatial coordinates such that $r_1$ and $r_2$ are both much 
larger than
$2GM/c^2$; then, neglecting higher-order multipole moments we find that \footnote{The general expression for the gravitomagnetic time delay in the gravitational field of higher-spin multipole moments of a stationary rotating body was derived in \cite{[3a]}, equation (43). A sign error in formula (43) must be corrected: the overall sign of this expression must be reversed. }
\begin{equation}
\Delta_{GM} = -\frac{2G}{c^4}\;\int^{P_2}_{P_1}\;\frac{{\bf J}\cdot({\bf 
r}\times d{\bf
r})}{r^3}\;\;.
\end{equation}
A detailed, but straightforward, calculation shows that 
\begin{equation}\label{16}
\Delta_{GM}=-\frac{2GJ}{c^4}\;\left(\frac{1}{r_1} +
\frac{1}{r_2}\right)\;\;\frac{{\bf\hat J}\cdot({\bf\hat r}_1 
\times{\bf\hat
r}_2)}{1+{\bf\hat r}_1\cdot {\bf\hat r}_2}\;\;.
\end{equation}
Thus $\Delta_{GM}$ also vanishes in this case if ${\bf J}, {\bf 
r}_1$ and
${\bf r}_2$ are in the same plane. 

Let us note that equation (\ref{16}) is exact and holds in general for the exterior of any spherically symmetric slowly rotating mass with constant ${\bf J}$. Moreover, let $d$ be the impact parameter of the light ray; then, when $r_1/d>> 1$ and $r_2/d>> 1$ it is possible to show that $\Delta_{GM}\simeq -4G{\bf J}\cdot{\bf\hat n}/(c^4d)$, where ${\bf\hat n}$ is a unit vector normal to the plane formed by ${\bf r}_1$ and ${\bf r}_2$ and directed along ${\bf r}_1\times{\bf r}_2$. Extending this calculation to the case in which the source is a thin rotating spherical shell of constant radius $R_0$ and the light ray penetrates the shell such that the portion of the ray inside the shell subtends an angle $\varphi$ at the center of the shell, we find from equations (\ref{14}) and (\ref{16}) that for $r_1>>R_0$ and $r_2>>R_0$,   $\Delta_{GM}\simeq-4G{\bf J}\cdot{\bf\hat n}f(\varphi)/(c^4d)$, where $f(\varphi)=\tan\left({\pi-\varphi\over 4}\right)+{1\over 2}\sin\varphi$ varies from $f(0)=1$ to $f(\pi)=0$ as $\varphi: 0\rightarrow\pi$. The impact parameter of the ray is given by $d=R_0\cos{\varphi\over 2}$, so that for $\varphi=0$ we recover the previous expression for $\Delta_{GM}$.

In two recent papers \cite{[14]}, \cite{[15]},  
the gravitomagnetic time delay in different images due to
gravitational lensing has been studied and a derivation of the time delay
due to the spin of an external shell has been presented. These results, together
with a discussion of the separability of 
time delays of different origin in the images, 
demonstrate that the
gravitomagnetic time delay would in principle be measurable in the near future.  
In particular, the
gravitomagnetic time delay must be taken into account in the
analysis of gravitational lensing time delay of some
extragalactic sources, since the estimated $\Delta_{GM}$ may
exceed the present measurement uncertainty of $\sim 0.5$ day in
the lensing time delay; we discuss this issue in section \ref{sec:5}.  Further measurement aspects of
$\Delta_{GM}$ are explored in \cite{[16]}, \cite{[17]}.  It is hoped that the
measurement of $\Delta_{GM}$ might provide information about the
existence and distribution of rotating dark matter.

It would be interesting to estimate how much of the gravitational time 
delay between
any two local points $P_1$ and $P_2$ could be due to the cosmological content of the
universe.  To this end, we consider the development of a Fermi coordinate 
system in the
neighborhood of a fundamental geodesic observer in the standard FLRW model 
\cite{[18]}.  At
an epoch with Hubble ``constant'' $H$, the gravitational time delay over 
any local
distance $r$ compared to $r/c$ is of the order $H^2 r^2/c^2$, which is 
negligibly
small.  A similar estimate holds for a rotating universe model as 
explained in the next
section.

\section{G\"odel universe}\label{sec:4}

The metric of the stationary and spatially homogeneous G\"{o}del universe 
can be
expressed as \cite{[19]}
\begin{equation} ds^2 = -dt^2-2\sqrt{2}\; U(x)\;dtdy + dx^2 - U^2(x) dy^2 
+ dz^2\;\;,
\end{equation}
where $U(x) ={\rm exp}\,(\sqrt{2}\;\Omega x)$ and we use units 
such that $c=1$
unless specified otherwise.  It turns out that for this metric the Ricci 
curvature is
given by $R_{\mu\nu}=2\Omega^2\,u_{\mu} u_{\nu}$, where 
$u^{\mu}=\delta^{\mu}_{\;\;0}$
is the velocity vector of a particle at rest in space and coincides with 
the timelike
Killing vector $\partial_t$.  The physical content of this model may be 
thought of as a
perfect fluid with velocity $u^{\mu}$ and constant density and pressure 
given by
$\rho=p=\Omega^2/(8\pi G)$, where $\Omega\partial_z$ is the vorticity vector associated
with the geodesic worldlines of the fluid.  Alternatively, the universe 
could be filled
with dust of constant density $\Omega^2/(4\pi G)$ together with a 
cosmological constant
$\Lambda = -\Omega^2$ \cite{[20]}.

To study the influence of this rotating cosmos on the local physics, we 
establish a
Fermi normal coordinate system in the neighborhood of a standard observer 
in this
model.  Imagine therefore an observer at rest in space and comoving with 
the fluid;
then, the observer follows a geodesic and its velocity vector $u^{\mu} =
\lambda^{\mu}_{\;\;(0)}$ is the temporal axis of an orthonormal tetrad 
frame
$\lambda^{\mu}_{\;\;(\alpha)}$ that is parallel transported along the geodesic.  In
$(t, x, y, z)$ coordinates, such a tetrad is given by
\begin{eqnarray}
\lambda^{\mu}_{\;\;(0)} &=& (1, 0, 0, 0)\;\;\;,\\
\lambda^{\mu}_{\;\;(1)} &=& \tilde{\lambda}^{\mu}_{\;\;(1)}\;{\rm 
cos}\;\Omega t +
\tilde{\lambda}^{\mu}_{\;\;(2)}\;{\rm sin}\;\Omega t\;\;,\\
\lambda^{\mu}_{\;\;(2)} &=& -\tilde{\lambda}^{\mu}_{\;\;(1)}\;{\rm 
sin}\;\Omega t +
\tilde{\lambda}^{\mu}_{\;\;(2)}\;{\rm cos}\;\Omega t\;\;,\\
\lambda^{\mu}_{\;\;(3)} &=& (0, 0, 0, 1)\;\;\;,
\end{eqnarray}
where $\tilde{\lambda}^{\mu}_{\;\;(1)}$ and 
$\tilde{\lambda}^{\mu}_{\;\;(2)}$
are defined by
\begin{eqnarray}
\tilde\lambda^{\mu}_{\;\;(1)} &=& (0, 1, 0, 0)\;\;\;,\\
\tilde\lambda^{\mu}_{\;\;(2)} &=& (-\sqrt{2}, 0, U^{-1}(x), 0)\;\;.
\end{eqnarray}
Here the rotation of the orthonormal triad
$\lambda^{\mu}_{\;\;(i)}, i = 1,2,3$, representing ideal
gyroscope directions characterizing the spatial Fermi frame,
about the $z$-axis with frequency $\Omega$ has been made
explicit.  The Fermi coordinate system assigns coordinates
$X^{\mu} = (T, X, Y, Z)$ to an event $P$ in the neighborhood of
the geodesic worldline under consideration as follows : there
exists a unique spacelike geodesic connecting $P$ to the
worldline at $P_0$ such that the two paths are orthogonal at
$P_0$, i.e. $\lambda^{\mu}_{\;\;(0)}\;\eta_{\mu} = 0$, where
$\eta_{\mu}$ is the unit tangent vector to the spacelike geodesic
at $P_0$.  Let $\tau$ be the proper time of the observer at $P_0,
\sigma$ be the proper length of the spacelike geodesic segment
$P_0 P$ and $C_i = \eta_{\mu}\lambda^{\mu}_{\;\;(i)}$ be the
direction cosines of this segment with respect to the observer's
triad at $P_0$. Then $T = \tau, X = \sigma C_1, Y = \sigma C_2$
and $Z=\sigma C_3$.  In these Fermi coordinates, the spacetime in
the neighborhood of the observer is Minkowskian except for the
cosmic tidal influence expressed by the gravitational potential
$^{F}h_{\mu\nu}$, i.e. $g_{\mu\nu} = \eta_{\mu\nu} +
{^{F}h}_{\mu\nu}$, where
\begin{eqnarray} ^{F}h_{00} &=& - ^{F}R_{0i0j} X^iX^j \;\;,\\
^{F}h_{0i} &=& -\frac{2}{3}\;\;{^{F}R}_{0jik}X^{j}X^k\;\;,\\
^{F}h_{ij} &=& -\frac{1}{3}\;\;{^{F}R}_{ikjl}X^k X^l \;\;.
\end{eqnarray}
Here the Riemann curvature components are the projections of the 
Riemann
tensor on the tetrad of the observer
\begin{equation}
^{F}R_{\alpha\beta\gamma\delta}=R_{\mu\nu\rho\sigma}\lambda^{\mu}_{\;\;(\alpha)}\;
\lambda^{\nu}_{\;\;(\beta)}\lambda^{\rho}_{\;\;(\gamma)}\lambda^{\sigma}_{\;\;(\delta)}\;\;.
\end{equation}

It turns out that all of the nonzero components of the Riemann tensor in 
$x
^{\mu}=(t, x, y, z)$ coordinates can be obtained from
\begin{eqnarray}
R_{0101} &=& \Omega^2\;\;,\;\;R_{0202}\;=\;\Omega^2 U^2(x)\;\;,\\
R_{0112} &=& -\sqrt{2}\;\Omega^2U(x)\;\;,\;\;R_{1212} = 3\;\Omega^2 U^2 
(x)\;\;,
\end{eqnarray}
using the symmetries of the Riemann tensor.  It follows that in 
the Fermi
coordinates all of the nonzero components are given by
\begin{equation}
{^FR}_{0101} = {^FR}_{0202} = {^FR}_{1212} = \Omega^2
\end{equation}
via the symmetries of the Riemann tensor.  Thus
\begin{eqnarray}
{^Fh}_{00} &=& -\Omega^2(X^2 + Y^2)\;\;,\;\;{^Fh}_{11} = 
-\frac{1}{3}\Omega^2
Y^2\;\;,\\
{^Fh}_{12} &=& {^Fh}_{21} = \frac{1}{3}\Omega^2
XY\;\;,\;\;{^Fh}_{22}=-\frac{1}{3}\Omega^2X^2\;\;,
\end{eqnarray}
are the only nonzero components of the gravitational potential 
and the metric
at this linear order is given by
\begin{eqnarray}\nonumber
ds^2 = &-&\left[1+\Omega^2(X^2+Y^2)\right]dT^2 + dX^2 + dY^2 + dZ^2 
\\\nonumber\\&-&
\frac{1}{3}\Omega^2(XdY - YdX)^2\;\;.
\end{eqnarray}

It is now straightforward to compute the gravitational time delay (4) using the
relations (31) - (32).  The result is
\begin{eqnarray}
\Delta_{\rm G\ddot{o}del} = &-&\frac{1}{6c^3}\Biggl\{
\frac{[\mbox{\boldmath$\Omega$}\cdot({\bf R}_1 \times {\bf R}_2)]^2}{|{\bf 
R}_1 - {\bf
R}_2|} + |{\bf R}_1 - {\bf R}_2| [(\mbox{\boldmath$\Omega$}\times{\bf 
R}_1)^2
\nonumber\\\nonumber\\  &+& (\mbox{\boldmath$\Omega$}\times{\bf R}_2)^2 +
(\mbox{\boldmath$\Omega$} \times {\bf R}_1)\cdot 
(\mbox{\boldmath$\Omega$}\times {\bf
R}_2)]\Biggr\}\;\;,
\end{eqnarray}
where $\mbox{\boldmath$\Omega$} = \Omega{\bf\hat Z}$ and ${\bf 
R}_1$ and
${\bf R}_2$ are the position vectors of $P_1$ and $P_2$, respectively.

The relative time delay for a signal from the observer $({\bf R}_1 = 0)$ 
to a local
distance $R$ compared to $R/c$ is thus given by $(\mbox{\boldmath$\Omega$} 
\times
{\bf R})^2/c^2$, which is negligibly small on the basis of the current 
upper limit of
$\Omega\leq 10^{-24}$ sec$^{-1}$ on the possible rate of rotation of the 
universe
\cite{[2]}, \cite{[21]}--\cite{[24]}.

\section{Gravitomagnetic time delay and gravitational lensing}\label{sec:5}

In gravitational lensing, depending upon
the relative position and distance
of the source, observer and deflecting mass, several images of the
same source may be observed. Numerous examples of gravitational lensing
have been discovered; a well-known case is the gravitational lens
Q2237 + 0305, or Einstein Cross, where the path of light from a
quasar estimated to be at a distance of approximately 8 billion
light years is bent by the gravitational field of a galaxy
estimated to be at a distance of about 400 million light years.
This light bending produces four images of the same quasar as
observed from the Earth \cite{[26]}, \cite{[27]}.

The gravitomagnetic time delay due to the spin of an astrophysical
object might then be detected in different images of the same
source by gravitational lensing.
Time delay in the images may be generated by the spins of the
deflecting object and of other bodies around the path of the light
rays, e.g. by the spin of an external rotating mass.

Let us estimate the spin time delay corresponding to some
astrophysical configurations. For simplicity, to derive the order
of magnitude of the gravitomagnetic time delay we assume that the
source, lens and observer are aligned. Nevertheless, there is an
additional time delay, called geometrical time delay \cite{[28]}, due to
the different geometrical paths followed by different rays.
Depending on the geometry of the system, this additional term may
be very large and may be the main source of time delay. However,
if we compare the time delay of photons that follow the same
geometrical path we can neglect the geometrical time delay, as in
the case of two light rays with the same impact parameter but on
different sides of the deflecting object. For a small deflection
angle, the contribution to the travel time delay from the
deflected path length traveled is
of the second order in the deflecting potentials; however, depending on the
geometrical configuration considered, this delay may need to be
included in the total time delay. Here, for simplicity, we
neglect any geometrical time delay. Furthermore, there is a
relative time delay due to the quadrupole moment of the central
mass distribution. It is shown in \cite{[14]} how, in the special case
of multiple images of the source with the same impact parameter,
propagating along the same axis, one may in principle have enough
observables to solve for the angular momentum $J$ and mass
$M$ of the central deflecting body by eliminating the time delay
due to the unknown quadrupole moment. Of course, for other
configurations in which the source is not exactly aligned with the 
lens and the observer, the difference in the paths traveled and the
corresponding difference in the Shapiro time delay can be the main
source of relative time delay; one would then need to model and
remove these delays between the different images on the basis of
the observed geometry of the system.

Let us now calculate the time delay due to the spin of some
astrophysical sources. For the Sun, $GM_\odot/c^2 \simeq 1.477$ km and
$R_\odot \simeq 6.96\times 10^5$ km, by considering two light rays
traveling on the equatorial plane with impact parameters $b \simeq
R_\odot$ and $- b$, the relative gravitomagnetic time delay from
formula (16) is given by $\Delta_{GM}^{\rm rel}=8GJ/(c^4b)\simeq1.54 \times
10^{-11}$ sec. The time delay due to spin of the Sun could then, in
principle, be measured using an interferometer at a distance of
about $8\times 10^{10}$ km by detecting the gravitationally deflected
photons emitted by a laser on the side of the Sun opposite the
detector and traveling on opposite sides of the Sun to the interferometer. To derive
the time delay due to the lensing galaxy of the Einstein Cross
\cite{[26]}, \cite{[27]}, we assume a simple model for the rotation and shape of the
central object. Details about this model can be found in \cite{[29]}.
The angular separation between the four images is about $0.9''$,
corresponding to a radius of closest approach of about $
650 h^{-1}_{75}$ pc, and the mass inside a shell with this radius
is $\sim 1.4\times 10^{10}\,h^{-1}_{75}\,M_{\odot}$ \cite{[27]}. Let
us assume that  $GJ/c^3\simeq 10^{23}h^{-2}_{75}$ km$^2$, we then have
from formula (16) that $\Delta_{GM}^{\rm rel}=8GJ/(c^4b)\simeq 4$ min.
Thus, at least in principle, one could measure the time delay due
to the spin of the lensing galaxy; of course, as in the case of
the Sun, one should be able to model with sufficient accuracy and remove
all the other delays due to other physical effects from the
observed time delays between the images. As a third example we
consider the relative time delay of photons due to the spin of a
typical cluster of galaxies of mass
$M_C \simeq 10^{14} M_{\odot}\,$, radius $R_C \simeq 5 \,$ Mpc and
angular velocity $\omega_C \simeq 10^{-18}$ sec$^{-1}$; depending
on the geometry of the system and on the path followed by the
photons, we then find relative time delays ranging from a few
minutes to several days \cite{[14]}.

Especially interesting is the case of gravitomagnetic time delay
due to the rotation of an external sphere. Let us now calculate
the time delay in the travel time of photons propagating inside a
rotating shell corresponding to some astrophysical
configurations. In the case of the Einstein Cross \cite{[26]}, in
order to get an order of magnitude of the effect, we assume that
the lensing galaxy has a radius $R\simeq 5$ kpc; after
some calculations based on the model given in \cite{[29]} and
integrating formula (14), the relative time delay of two photons
traveling at distances of $650$ pc on opposite sides from the center is given by
$\Delta_{GM}^{\rm rel}\simeq 20$ min. If the lensing galaxy is inside
a rotating cluster, or super-cluster of galaxies, to get an order
of magnitude of the time delay due to the spin of the mass
rotating around the deflecting galaxy, we use typical
super-cluster parameters: total mass $M=10^{15}\,M_\odot$, radius
$R=70$ Mpc and angular velocity $\omega=2 \times 10^{-18}$ sec$^{-1}$ \cite{[30]}. If the galaxy is in the center of the cluster and light
rays have impact parameters $\simeq 15$ kpc (of the order of the Milky Way radius), the time
delay, in the case of constant density $\rho$ can be determined by
integrating formula (14) and the result is $\Delta_{GM}^{\rm rel}\simeq 1$ day.

Finally, if the lensing galaxy is not at the center of the
cluster of radius $R$ but at a distance $a R$ from the center, with $0 \leq
a \leq 1$, by integrating formula
(14) between the impact parameter $r\simeq a R$ and $R$, we find that when the impact parameters $r_1$ and $r_2$ of two light rays are approximately equal
\begin{equation}
\Delta_{GM}^{\rm rel} = \frac{8 GM\omega}{3c^4}({r_1  - r_2}) 
{(1 - a^2)^{1/2}} (1 - 4 a^2) \;.
\end{equation}
Thus, if the lensing galaxy is at a distance of 10 Mpc from the
center of the cluster, the relative time delay due to the spin of
the external rotating mass between two photons with
$r_1-r_2\simeq 30$ kpc is $\Delta_{GM}^{\rm rel}\simeq 0.9$
day.

Promising candidates to observe the time delay due to spin are
systems of the type of the gravitational lens B0218+357 \cite{add-ref},
where the separation angle between the images is so small,
335 milliarcsec for B0218+357, that the time delay
between the images is also very small, about 10.5 days for the two
images of this system. In such cases, the total time delay is
comparable, for a favorable configuration,
to the time delay due to spin.
In addition, an Einstein ring is observed in the system B0218+357 and
the diameter of the ring is the same as the separation of the
images. In such a configuration, the Einstein ring can provide strong
constraints on the mass distribution in the lens; in turn, this
can be extremely useful in order to distinguish time delays
of different origin. 
Since the present measurement uncertainty in the lensing time
delay is of the order of 0.5 day, the gravitomagnetic time delay
might already be observable \cite{[31]}.
 For example in the system
B0218+357 the measured delay is 10.5$\pm$0.4 day.
This measurement is possible because the source is a
strongly variable radio object, thus one can determine the time
delay of the variations in the images. In the case of B0218+357
it is possible to observe distinct variations in the total flux
density, percentage polarization and
the polarization position angle at two frequencies.

\section{Discussion}\label{sec:6}

Time delay plays an important role in classical wave scattering
as well as quantum scattering theory (see \cite{[25]} for a recent
comprehensive survey of this subject).  Here we have confined our
attention to the propagation of radiation in the JWKB limit, i.e.
null rays, in a gravitational field that is considered only to
linear order.  We have then derived expressions for the
gravitomagnetic time delay of null rays propagating inside a
spinning shell and in the field of rotating masses.

Since we have shown that there may be an appreciable time delay
due to the spin of a body, or external shell, we draw the following 
conclusions.

The gravitomagnetic time delay should be taken into
account in the modeling of relative time delays of the images of
a source observed at a far point by gravitational lensing. This
effect is due to the propagation of the photons in opposite
directions with respect to the sense of the spin of the body,
or external rotating shell.

If other time delays can be modeled with sufficient accuracy and removed
from the observational data, the larger relative delay due to the
quadrupole moment of the lensing body can be removed, at least in
principle and for some configurations of the images, by using
special combinations of the observables; thus, one could directly
measure the spin time delay due to the gravitomagnetic field of
the lensing body. In order to estimate the relevance of the spin
time delay in some real astrophysical configurations, we have
considered some possible astrophysical cases. We have studied the
relative time delay in the gravitational lensing images caused by
a typical rotating galaxy or a cluster of galaxies. We have also
analyzed the relative gravitomagnetic time delay when the path of
photons is inside a galaxy, a cluster or a super-cluster of
galaxies rotating around the deflecting body; this effect should
be large enough to be detectable from the Earth.

The measurement of the gravitomagnetic time delay due to the
angular momentum of an external massive rotating sphere might
be a further observable for the determination of the total
mass-energy of the external body, i.e. of the dark matter content
of galaxies, clusters and super-clusters of galaxies. Indeed, by
measuring the gravitomagnetic time delay one can determine the
total angular momentum of the rotating body and thus, by
estimating the contribution of the visible part, one can
determine its dark-matter content.

These estimates are preliminary because we need to apply the
gravitomagnetic time delay to some particular, known,
gravitational-lensing images. Furthermore, we need to estimate
the size and the possibility of modeling of the other sources of time
delay. Nevertheless, depending on the geometry of the
astrophysical system considered, we conclude that the relative
gravitomagnetic time delay may be an already observable effect.

\newpage
\begin{figure*}
\centerline{\psfig{figure=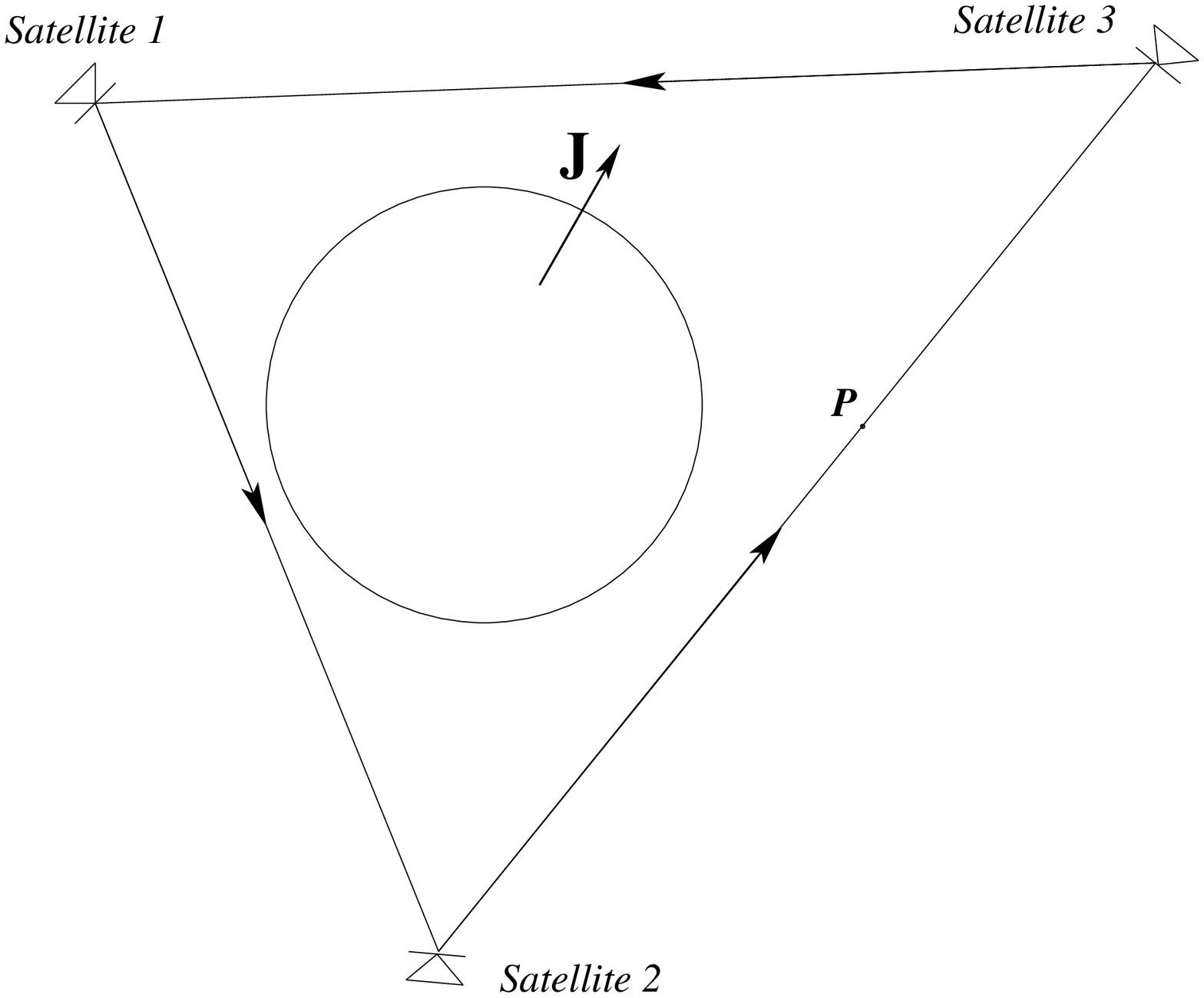,angle=0,height=19cm,width=17.5cm}} \caption{Light propagates from satellite 1, is transponded from satellites 2 and 3, and returns back to satellite 1 by making a closed counterclockwise loop. Gravitoelectric time delay $\Delta_{GE}$ and gravitomagnetic time delay $\Delta_{GM}$ are due to the gravitational field of the rotating mass $M$ possessing spin ${\bf J}$. Point $P$ is an arbitrary point on the loop. The``clock effect'' at the point $P$ is given by equation (\ref{11}) and also discussed in \cite{[7]} using a different theoretical approach.}
\end{figure*}
\end{document}